\documentclass[twocolumn,amsmath,amssymb,floatfix,superscriptaddress,prb]{revtex4-2} 
\usepackage{url,epsf}
\usepackage{comment}
\usepackage{appendix}
\usepackage{physics}
\usepackage{multirow}
\usepackage[colorlinks=true,linkcolor=blue]{hyperref}
\usepackage{amsmath,bm,graphicx,epsfig}
\usepackage{xcolor}
\usepackage[english]{babel}
\usepackage{amsfonts,stmaryrd,amssymb}
\usepackage{enumerate} 

\usepackage{csquotes}
\MakeOuterQuote{"}

\newcommand{\B}{\mathcal{B}}

\newcommand{\bY}{\mathbf{Y}}
\newcommand{\bF}{\mathbf{F}}
\newcommand{\bA}{\mathbf{A}}

\newcommand{\be}{\begin{equation}}
\newcommand{\ee}{\end{equation}}
\newenvironment{eqs}
{\begin{equation} \begin{aligned}}
{\end{aligned} \end{equation} }
\newcommand{\bal}{\begin{eqs}}
\newcommand{\eal}{\end{eqs}}

\newcommand{\R}{\mathcal{R}}

\newcommand{\Lag}{\mathcal{L}}

\newcommand{\N}{\mathcal{N}}

\newcommand{\F}{\mathcal{F}}

\newcommand{\h}{\hat{H}}

\newcommand{\D}{\mathcal{D}}

\newcommand{\fa}{f^{\phantom{\dagger}}}
\newcommand{\fc}{f^\dagger}

\newcommand{\bea}{\begin{eqnarray}}
\newcommand{\eea}{\end{eqnarray}}
\newcommand{\ba}{\begin{eqnarray*}}
\newcommand{\ea}{\end{eqnarray*}}
\newcommand{\dagga}{{\phantom{\dagger}}}

\newcommand{\Av}[2]{\langle #1|\,#2\,|#1\rangle}

\begin{document}

\begin{abstract}

We introduce the time-dependent ghost Gutzwiller approximation (td-gGA), a non-equilibrium extension of the ghost Gutzwiller approximation (gGA), a powerful variational approach which systematically improves on the standard Gutzwiller method by including auxiliary degrees of freedom.

We demonstrate the effectiveness of td-gGA by studying the quench dynamics of the single-band Hubbard model as a function of the number of auxiliary parameters. 
Our results show that td-gGA captures the relaxation of local observables, in contrast with the time-dependent Gutzwiller method. This systematic and qualitative improvement leads to an accuracy comparable with time-dependent Dynamical Mean-Field Theory which comes at a much lower computational cost.

These findings suggest that td-gGA has the potential to enable extensive and accurate theoretical investigations of  multi-orbital correlated electron systems in nonequilibrium situations, with potential applications in the field of quantum control, Mott solar cells, and other areas where an accurate account of the non-equilibrium properties of strongly interacting quantum systems is required.
\end{abstract}

\title{Time-dependent ghost-Gutzwiller non-equilibrium dynamics}


\author{Daniele Guerci}
\affiliation{Center for Computational Quantum Physics, Flatiron Institute, New York, New York 10010, USA}
\author{Massimo Capone}
\affiliation{Scuola Internazionale Superiore di Studi Avanzati (SISSA), Via Bonomea 265, 34136 Trieste, Italy}
\affiliation{CNR-IOM, Istituto Officina dei Materiali, Consiglio Nazionale delle Ricerche, Via Bonomea 265, 34136 Trieste, Italy}

\author{Nicola Lanat\`a}
\altaffiliation{Corresponding author: nxlsps@rit.edu}
\affiliation{School of Physics and Astronomy, Rochester Institute of Technology, 84 Lomb Memorial Drive, Rochester, New York 14623, USA}
\affiliation{Center for Computational Quantum Physics, Flatiron Institute, New York, New York 10010, USA}

\date{\today}

\maketitle



\section{Introduction}

The study of the nonequilibrium dynamics of correlated electron systems has gained significant attention in recent years, ranging from fundamental questions~\cite{RevModPhys_Silva} to topics triggered by the emergence of experimental techniques that allow for the investigation of these systems under a variety of conditions. Examples include ultrafast spectroscopy techniques, which allow for the investigation of the dynamics of solid state materials on the femtosecond time scale~\cite{Cavalleri_2001,Giannetti_2016,Sentef_2021}
and the use of ultracold atoms in optical lattices~\cite{Jaksch_2005,RevModPhys.80.885,Esslinger_2010,Bloch2012Quantum,Mazurenko_2017,Drewes_2017}, which allows for the study of correlated quantum systems in a controlled and tunable environment. 
The development of efficient photovoltaic technologies such as Mott solar cells~\cite{Manousakis_2010,PhysRevB.90.235102,PhysRevApplied.3.064015,Petocchi_2019,Petocchi_2022}, which exploit the peculiar properties of correlated electron systems, is another example of the many directions calling for theoretical tools and frameworks able to investigate quantum many-body systems out of equilibrium.

The paradigmatic model in the study of correlated electron systems is the (single-orbital) Hubbard model, which describes fermionic particles on a lattice experiencing the effects of local repulsive interactions. A huge body of work has focused on the equilibrium properties of this model, while the investigation of the non-equilibrium physics is severely limited by technical aspects. 
Current state-of-the-art methods, such time-dependent Dynamical Mean-Field Theory (td-DMFT)~\cite{Rev_tDMFT,Eckstein_2009,Eckstein_2010,ReviewDMFT}, can be indeed computationally demanding for many applications and they are typically limited to relatively short time scales. This situation calls for the development of computationally lighter, yet sufficiently accurate, methods to study the dynamics of the Hubbard model and possibly of more involved and richer models. 

To address this challenge, here we exploit the so-called ghost Gutzwiller approximation (gGA)~\cite{Lanata-ghost-2017,PhysRevB.104.L081103,Daniele2019}, that generalizes the standard Gutzwiller approximation (GA)~\cite{Gutzwiller-2,NicolaPRX} systematically extending the variational space introducing auxiliary degrees of freedom. 
This perspective introduces similarities between this variational wave function and matrix-product states or more recent neural network states~\cite{Robledo_Moreno_2022} where the number of "hidden" degrees of freedom is directly connected with the amount of entanglement in the variational wave function.
Recently, the method has been formulated also in terms of a formally-exact rotationally-invariant slave boson theory (RISB)~\cite{lanata2021operatorial,NicolaRISB,Kotliar&Ruckenstein,PhysRevB.76.155102}, that reduces to the gGA within the mean-field approximation.
In equilibrium, the addition of $\B$ subsidiary fermionic degrees of freedom improves consistently the accuracy of the wave function, and allows for a faithful description of the Mott insulator which reproduces the main results of DMFT already for small values of $\B \lesssim 7$~\cite{Lee_2022}.

In this work we introduce a non-equilibrium extension of the gGA framework, that generalizes the standard time-dependent Gutzwiller approximation~\cite{Fabrizio_review,SchiroFabrizioPRL,Fabrizio_2012,Lanata_TD,LorenzanaPRL,Seibold2011a,PhysRevB.86.161101,Sandri2012,Sandri2013,PhysRevB.96.201115,Giacomo_2017,Guerci_2019}.
We apply the method to an interaction quench in the half-filled Hubbard model, a topic which attracted considerable interest, both in experiments~\cite{J_rdens_2008,schneider2012fermionic,schafer2020tools} and theoretical investigations~\cite{Rev_tDMFT,Moeckel_2008,Eckstein_2009,Eckstein_2010,PhysRevB.85.085129,ReviewDMFT}.

We show that the improvement introduced by td-gGA is substantial and qualitative. In particular the new method captures the relaxation of local observables, a crucial feature which is not accessible by the standard time-dependent Gutzwiller approximation~\cite{Fabrizio_review}. 

In addition, we show that, using a small number of auxiliary  variables, the td-gGA reproduces the dynamics obtained within td-DMFT for reasonably large timescale with a significantly reduced computational cost. As we shall describe in the following, the td-gGA requires indeed to solve a set of non-linear differential equations as opposed to the integro-differential equations obtained within td-DMFT.

%
Our results highlight the potential of td-gGA to substantially reduce the computational cost of accurate studies of the time-resolved dynamics of strongly correlated systems. This opportunity can open the path to effective investigations of multi-orbital correlated electron systems in nonequilibrium situations, extending the scope to a number of different correlated materials and enabling a variety of applications ranging from energy-related materials~\cite{Manousakis_2010,PhysRevB.90.235102,PhysRevApplied.3.064015,Petocchi_2019,Petocchi_2022} to quantum control~\cite{Qcontrol_lectures_2020}, and other areas where the accurate treatment of strong correlations is required.

 The plan of the paper is as follows. In Sec.~\ref{sec:method} we introduce the model and we formulate the time-dependent ghost-Gutzwiller approach, while in Sec.~\ref{sec:results} we apply the tools to study the dynamics of the single-band Hubbard model. The comparison with DMFT results that we thus obtain is discussed in Sec.~\ref{sec:results}. Finally, Sec.~\ref{sec:conclusions} is devoted to concluding remarks.

\section{Model and method}
\label{sec:method}

We consider the time-dependent dynamics of the single band Hubbard model at half filling: 
\begin{equation}
\label{Hubbard_model}
    \hat{H}=
    \frac{U}{2}\sum_i\left(\hat{n}_i-1\right)^2
    -J\sum_{\langle i,j \rangle}\sum_{\sigma=\uparrow,\downarrow}\left(c^\dagger_{i\sigma}c_{j\sigma}+H.c.\right),
\end{equation}
where $U$ is the Hubbard on-site interaction strength, $\hat{n}_i=\sum_{\sigma=\uparrow,\downarrow} c^\dagger_{i\sigma}c_{i\sigma}$ is the local occupancy operator,
$J$ the hopping between nearest-neighbor sites. We consider the model on a Bethe lattice with semicircular density of states $\rho(\omega)=2\sqrt{D^2-\epsilon^2}/(\pi D^2)$ and we measure energy in unit of the half-bandwidth $D \propto J$. 
From now on we set $D$ as energy unit and $D^{-1}$ as time unit.

In this work we focus on the time-resolved evolution of the system in a popular non-equilibrium protocol, the interaction quench, where the interaction is suddenly changed from $U_i$ to $U_f$. As a matter of fact, we will evaluate the time evolution governed by Eq.~\eqref{Hubbard_model} for $U=U_f$ using as an initial state the equilibrium solution for $U=U_i$.

\subsection{Equlibrium gGA Lagrange function}

Specializing the formalism of Refs.~\cite{Lanata-ghost-2017,PhysRevB.104.L081103}, to the single-orbital model Eq.~\eqref{Hubbard_model} and enforcing spin rotational invariance and translational invariance, we obtain that the gGA ground state is encoded in the following Lagrange function:
\begin{align}
&\Lag[
{\Phi},E^c;\,  \R,\Lambda;\, \D, \Lambda^{c};\,\Delta, \Psi_0, E]=
\nonumber\\&\;
=\frac{1}{\mathcal{N}}\Av{\Psi_0}{\h_{\text{qp}}[\R,\Lambda]}
+E\!\left(1\!-\!\langle\Psi_0|\Psi_0\rangle\right)
\nonumber\\&\;
+\left[\Av{\Phi}{\h_{\text{emb}}[\D,\Lambda^c]}
+E^c\!\left(1-\langle \Phi | \Phi \rangle
\right)\right]
\nonumber\\&\,
-\left[
\sum_{\sigma=\uparrow,\downarrow}\sum_{a,b=1}^{\B}\big(
\Lambda_{ab}+\Lambda^c_{ab}\big)\Delta_{ab}
\right.
\nonumber\\&\left.
\qquad+\sum_{\sigma=\uparrow,\downarrow}
\sum_{c,a=1}^{\B}
\big(
\D_{a} \R_{c}
\left[\Delta(1-\Delta)\right]^{\frac{1}{2}}_{ca}
+\text{c.c.}\big)
\right]\,,
\label{Lag-SB-emb}
\end{align}
where $\mathcal{N}$ is the total number of unit cells,
$E$ and $E^c$ are real numbers, $\Delta$, $\Lambda^c$ and $\Lambda$ are
$\B\times \B$ Hermitian matrices, $\D$ and $\R$ are rectangular $\B\times 1$ matrices (whose row entries are $\D_a$ and $\R_a$, respectively).
The auxiliary Hamiltonians $\h_{\text{qp}}$ and $\h_{\text{emb}}$,
which are called "quasiparticle Hamiltonian" and "Embedding Hamiltonian" (EH),
respectively, are defined as follows:
\begin{align}
    \h_{\text{qp}}&=-J\sum_{\langle i,j \rangle}\sum_{a,b=1}^{\B}
    \sum_{\sigma=\uparrow,\downarrow}
    \R_a^\dagga \R^{\dagger}_b\,
    \fc_{ia\sigma}\fa_{jb\sigma}
    \nonumber\\&\quad 
    + \sum_{i}\sum_{a,b=1}^{\B}\sum_{\sigma=\uparrow,\downarrow} \Lambda_{ab}\fc_{ia\sigma} \fa_{ib\sigma}
    \label{hqp}
    \\
    &=\sum_{a,b=1}^{\B}\sum_{\omega=1}^\N \sum_{\sigma=\uparrow,\downarrow}
    \left(
    \epsilon_\omega\, \R_a^\dagga \R^{\dagger}_b
    + \Lambda_{ab}\right) \eta^\dagger_{\omega a\sigma}\eta^\dagga_{\omega b\sigma}
    \nonumber
    \\
    \h_{\text{emb}}&= 
    \frac{U}{2}\left(\hat{n}-1\right)^2
    + \sum_{a=1}^{\B}\sum_{\sigma=\uparrow,\downarrow}\left[\D_{a}\,\hat{c}^{\dagger}_{\sigma}\hat{f}^{\phantom{\dagger}}_{a\sigma}
    +\text{H.c.}\right]\nonumber\\
&\quad+\sum_{a,b=1}^{\B}\sum_{\sigma=\uparrow,\downarrow}\Lambda^c_{ab}\, \hat{f}^{\phantom{\dagger}}_{b\sigma}\hat{f}^{\dagger}_{a\sigma}
   \,,
   \label{hemb}
\end{align}
where $\epsilon_\omega$ are the eigenvalues of the hopping matrix for the Bethe lattice, $\eta_{\omega\sigma}$ 
are the corresponding eigenmodes, 
and $\hat{n}=\sum_{\sigma=\uparrow,\downarrow} \hat{c}^\dagger_{\sigma}\hat{c}_{\sigma}$ is the impurity occupancy operator.

The integer parameter $\B$ controls the size of the variational space and, in turn, the accuracy of the gGA solution.
In particular, for $\B=1$ Eq.~\eqref{Lag-SB-emb} reduces to the standard GA Lagrange function, while for higher values of $\B$ the accuracy of the gGA method is comparable to DMFT~\cite{Lanata-ghost-2017,PhysRevB.104.L081103}.

The saddle-point of the Lagrangian $\Lag$ defined in Eq.~\eqref{Lag-SB-emb} is given by the following equations:
\begin{align}
    &\int_{-D}^Dd\omega\,\rho(\omega)\,[n(\omega)]_{ab}
     = \Delta_{ab}
    \label{detDelta}
    \\
    &\int_{-D}^Dd\omega\,\rho(\omega)\,\omega
    \left[\R^{\dagger}\,^tn(\omega)\right]_{1 a}
    = \sum_{c,a=1}^{\B}\D_{c}\left[\Delta\left(1-\Delta\right)\right]^{\frac{1}{2}}_{ac}
    \label{detD}
    \\
   &\sum_{c,b=1}^{\B}\frac{\partial}{\partial d_s} \left(\left[\Delta\left(1-\Delta\right)\right]^{\frac{1}{2}}_{cb}\D_{b}\R_{c} + \mathrm{c.c.}\right) + [l+l^c]_{s} = 0
   \label{detLc}
   \\
   &\h^{\mathrm{emb}}\ket{\Phi} = E^c\ket{\Phi}
   \\
   &\F^{(1)}_{a} = \bra{\Phi}\hat{c}^{\dagger}_{\sigma}\hat{f}^{\phantom{\dagger}}_{a\sigma}\ket{\Phi}
   - \sum_{c=1}^\B\left[\Delta\left(1-\Delta\right)\right]_{ca}^{\frac{1}{2}}\R_{c} = 0
   \label{detF1}
   \\
   &\F^{(2)}_{ab} = \bra{\Phi}\hat{f}^{\phantom{\dagger}}_{b\sigma}\hat{f}^{\dagger}_{a\sigma}\ket{\Phi} - \Delta_{ab} = 0
   \,,
   \label{detF2}
\end{align}
where Eqs.~\eqref{detDelta} and \eqref{detD} are evaluated for the Bethe lattice at $\N\rightarrow \infty$ and the limit of infinite coordination number, $^t M$ indicates the transpose of a matrix $M$,
\begin{equation}
    [n(\omega)]_{ab}=\Av{\Psi_0}{\eta^\dagger_{\omega a\sigma}\eta^\dagga_{\omega b\sigma}}
    =[f\left(\R\omega\R^{\dagger}+\Lambda\right)]_{ba}
\end{equation}
is the quasiparticle ground-state single-particle density matrix in the Bethe lattice eigenmodes basis,
$f$ is the zero-temperature Fermi function and we expressed the matrices $\Delta$, $\Lambda$ and $\Lambda^c$ in terms of the following expansion with respect to an orthonormal
basis of Hermitian matrices $\left\{h_s\right\}$ (with respect to the canonical scalar product $(A, B) = \Tr \left[A^{\dagger}B\right]$):
\begin{align}
    \label{coeffDelta}
    \Delta =& \sum_{s=1}^{\B^2} d_s\, {}^t h_s \\
    \label{coeffL}
    \Lambda =& \sum_{s=1}^{\B^2} l_s \,h_s  \\
    \label{coeffLc}
    \Lambda^c =& \sum_{s=1}^{\B^2} l^c_s \, h_s \,,
\end{align}
where $d_s$, $l_s$ and $l^c_s$ are real-valued coefficients.

\subsection{Algorithmic structure of the gGA}
The equations \eqref{detDelta}-\eqref{detF2} can be solved numerically as follows:
(1) Starting from an initial guess for the entries  $\R$ and $\Lambda$, compute $\Delta$ from Eq.~\eqref{detDelta};
(2) Compute $\D$ using Eq.~\eqref{detD};
(3) Determine the coefficients $l^c_s$ from Eq.~\eqref{detLc} and construct the matrix $\Lambda^c$ from Eq.~\eqref{coeffLc};
(4) Construct $\h^{\mathrm{emb}}$ from Eq.~\eqref{hemb} and calculate its ground state $\ket{\Phi}$ within the subspace with $1+\B$ Fermions (i.e., at half filling);
(5) Compute $\F^{(1)}$ and $\F^{(2)}$ from Eqs.~\eqref{detF1} and \eqref{detF2}.
The parameters $(\R,\Lambda)$ such that Eqs.~\eqref{detF1} and \eqref{detF2} are satisfied are computed numerically.

\subsection{Time-dependent gGA Lagrange function}

As explained in Refs.~\cite{Lanata-ghost-2017,PhysRevB.104.L081103}, the equilibrium gGA Lagrange function and equations summarized above can be obtanied by applying the standard standard multi-orbital GA formulation of Ref.~\cite{NicolaPRX} within an enlarged Hilbert space, including additional auxiliary Fermionic degrees of freedom. Equivalently, the same equations can be derived from the standard multi-orbital formulation of Ref~\cite{NicolaRISB}, by introducing additional auxiliary Fermionic and Bosonic degrees of freedom~\cite{lanata2021operatorial}.
The td-gGA framework is straightforwardly obtained by applying the standard td-GA formalism of Ref.~\cite{Fabrizio_review}, simply by including such auxiliary degrees of freedom mentioned above, either from the GA perspective~\cite{guerci2019beyond} or, equivalently, from the RISB~\cite{lanata2021operatorial} perspective. The resulting dynamics is obtained by extremizing the following Lagrange function, previously introduced in Ref.~\cite{lanata2021operatorial}: 
\begin{align}
    \mathcal L&=\frac{1}{\N}\mel{\Psi_0}{i\partial_t-\h_{\rm qp}}{\Psi_0}+\mel{\Phi}{i\partial_t-\h_{\rm emb}}{\Phi}\nonumber\\
    &+\left[
\sum_{\sigma=\uparrow,\downarrow}\sum_{a,b=1}^{\B}
\Lambda^c_{ab}\Delta_{ab}
\right.
\nonumber\\&\;\left.
+\sum_{\sigma=\uparrow,\downarrow}
\sum_{c,a=1}^{\B}
\big(
\D_{a} \R_{c}
\left[\Delta(1-\Delta)\right]^{\frac{1}{2}}_{ca}
+\text{c.c.}\big)
\right]
\end{align}
where $\h_{\text{qp}}$ and $\h_{\text{emb}}$ are given by Eqs.~\eqref{hqp} and \eqref{hemb}, respectively, setting $\Lambda=0$.

As for the equilibrium case, the stationarity conditions with respect to $\Delta$, $\D$ and $\Lambda^c$
are Eqs.~\eqref{detLc}, \eqref{detF1}, \eqref{detF2}, respectively.
Instead, from the Dirac-Frenkel principle it follows that the stationarity condition with respect to $\ket{\Psi_0}$ and $\ket{\Phi}$ with the corresponding time-dependent Schr\"odinger equations. In summary, the dynamics of the gGA variational parameters is governed by the following equations:
\begin{align}
   &\left[i\partial_t-\h_{\text{emb}}\right]\ket{\Phi} = 0
   \label{dyn-Phi}
   \\
   & i\partial_t n_{ab}(\omega)=\omega \sum_{c=1}^\B\left(\R_b\R^\dagger_c \,n_{ac}(\omega)-\R_c\R^\dagger_a \,n_{cb}(\omega)\right)
   \label{dyn-nomega}
   \\
   &\int_{-D}^Dd\omega\,\rho(\omega)\,\omega
   \left[\R^{\dagger}\,^tn(\omega)\right]_{1 a}
    = \sum_{c,a=1}^{\B}\D_{c}\left[\Delta\left(1-\Delta\right)\right]^{\frac{1}{2}}_{ac}
    \label{tddetD}
    \\
   &\sum_{c,b=1}^{\B}\frac{\partial}{\partial d_s} \left(\left[\Delta\left(1-\Delta\right)\right]^{\frac{1}{2}}_{cb}\D_{b}\R_{c} + \mathrm{c.c.}\right) + l^c_{s} = 0
   \label{tddetLc}
   \\
   &\bra{\Phi}\hat{c}^{\dagger}_{\sigma}\hat{f}^{\phantom{\dagger}}_{a\sigma}\ket{\Phi}
   - \sum_{c=1}^\B\left[\Delta\left(1-\Delta\right)\right]_{ca}^{\frac{1}{2}}\R_{c} = 0
   \label{tddetF1}
   \\
   &\bra{\Phi}\hat{f}^{\phantom{\dagger}}_{b\sigma}\hat{f}^{\dagger}_{a\sigma}\ket{\Phi} - \Delta_{ab} = 0
   \,,
   \label{tddetF2}
\end{align}
where Eq.~\eqref{dyn-nomega} describes the time evolution of the quasiparticle single-particle density matrix of 
corresponding to the following time-dependent Schr\"odinger equation $
    \left[i\partial_t-\h_{\text{qp}}\right]\ket{\Psi_0} = 0$.

To implement the dynamics governed by Eqs.~\eqref{dyn-Phi}-\eqref{tddetF2} the integrals over $\omega$
are approximated by discretizing the interval $[-D,D]$ with a series of frequencies $\omega_n$.
The real and imaginary components of the vector $\Phi$ and of the matrices $n(\omega_n)$ are encoded into a  real-valued vector $\bY$.
Since $\R$, $\D$, $\Delta$ and $\Lambda^c$ can be all determined as a function of $\ket{\Phi}$ and $n(\omega)$ using 
Eqs.~\eqref{tddetLc}-\eqref{tddetF2}, and $\partial_t\ket{\Phi}$ and $\partial_t n(\omega)$ can be determined in terms
of these parameters using Eqs.~\eqref{dyn-Phi}-\eqref{dyn-nomega},
the dynamics of $\bY$ can be expressed as follows:
\begin{equation}
    \partial_t \bY(t)=\bF(\bY(t))\,,
\end{equation}
which is a non-linear first-order differential equation that can be integrated numerically with standard methods.
In particular, our calculations were performed using the
the Runge-Kutta library RKSUITE~\cite{brankin1993rksuite}.

It is important to note that the function $\bF$ used in our study of Hubbard quenches is independent of time when the Hamiltonian is not explicitly time-dependent after the sudden change in $U$. However, the equations we derived can still be used even if the Hamiltonian has an explicit time dependence. In this case, the td-gGA equations take the form: 
\begin{equation}
    \partial_t \bY(t)=\bF(\bY(t),\bA(t))\,,
    \label{eq-control}
\end{equation}
where $\bA(t)$ represents a time-varying external perturbation, such as an electromagnetic field.
This characteristic of the td-gGA framework is particularly interesting, because it opens the possibility of using standard frameworks, such as classical optimal control algorithms~\cite{Qcontrol_lectures_2020}, for manipulating the dynamics of electronic states.
In fact, such techniques are broadly applicable to dynamical systems governed by ordinary differential equations such as Eq.~\eqref{eq-control}, while
 frameworks for controlling the td-DMFT dynamics, governed by integro-differential equations, are not currently available.

\section{Interaction quench in the half-filled Hubbard model}
\label{sec:results}

\begin{figure}
    \centering
    \includegraphics[width=0.4\textwidth]{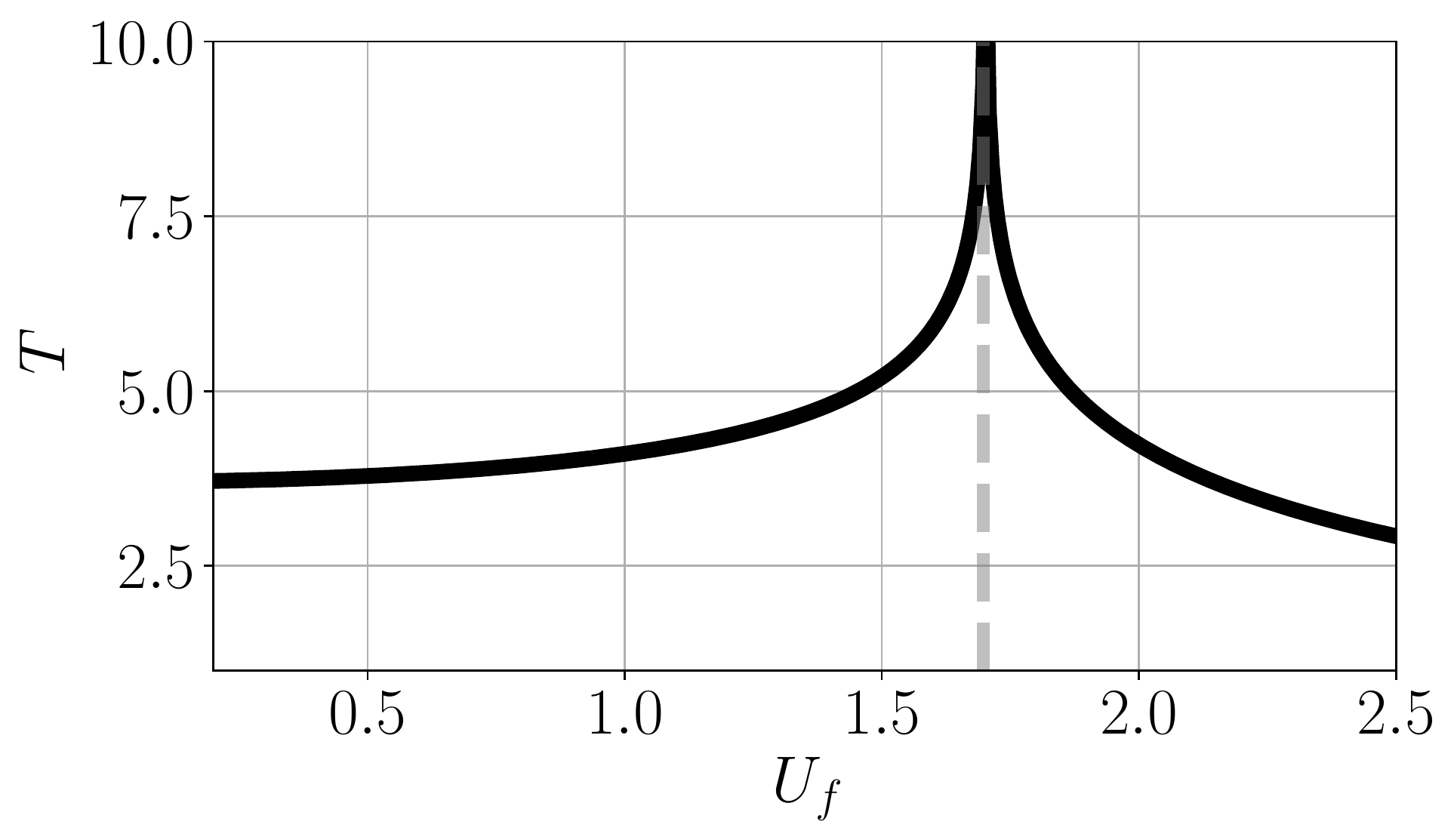}
    \caption{Period of the coherent oscillations in the dynamics of the single-band Hubbard model in the conventional Gutwiller approximation ($\B=1$) as a function of $U_f$. The divergence at $U^{\rm dyn}_c=U_c/2$ defined by the dashed grey vertical line shows the critical slowing down at the dynamical transition~\cite{SchiroFabrizioPRL,SchiroFabrizioPRB}. }
   \label{fig:periodo_standard}
\end{figure}

We now turn to the application of the formalism to discuss the out of equilibrium evolution in the half-filled single band Hubbard model~\ref{Hubbard_model}. 
The quantum quench protocol consists of preparing the system in the initial variational ground state of the model with interaction $U(t\le 0)=U_i$. Then, for $t> 0$, the state evolves under an Hamiltonian characterized by the value of the interaction $U(t>0)=U_f=U_i+\delta U\neq U_i$.
\begin{widetext}

\begin{figure}
    \centering
    \includegraphics[width=1.0\textwidth]{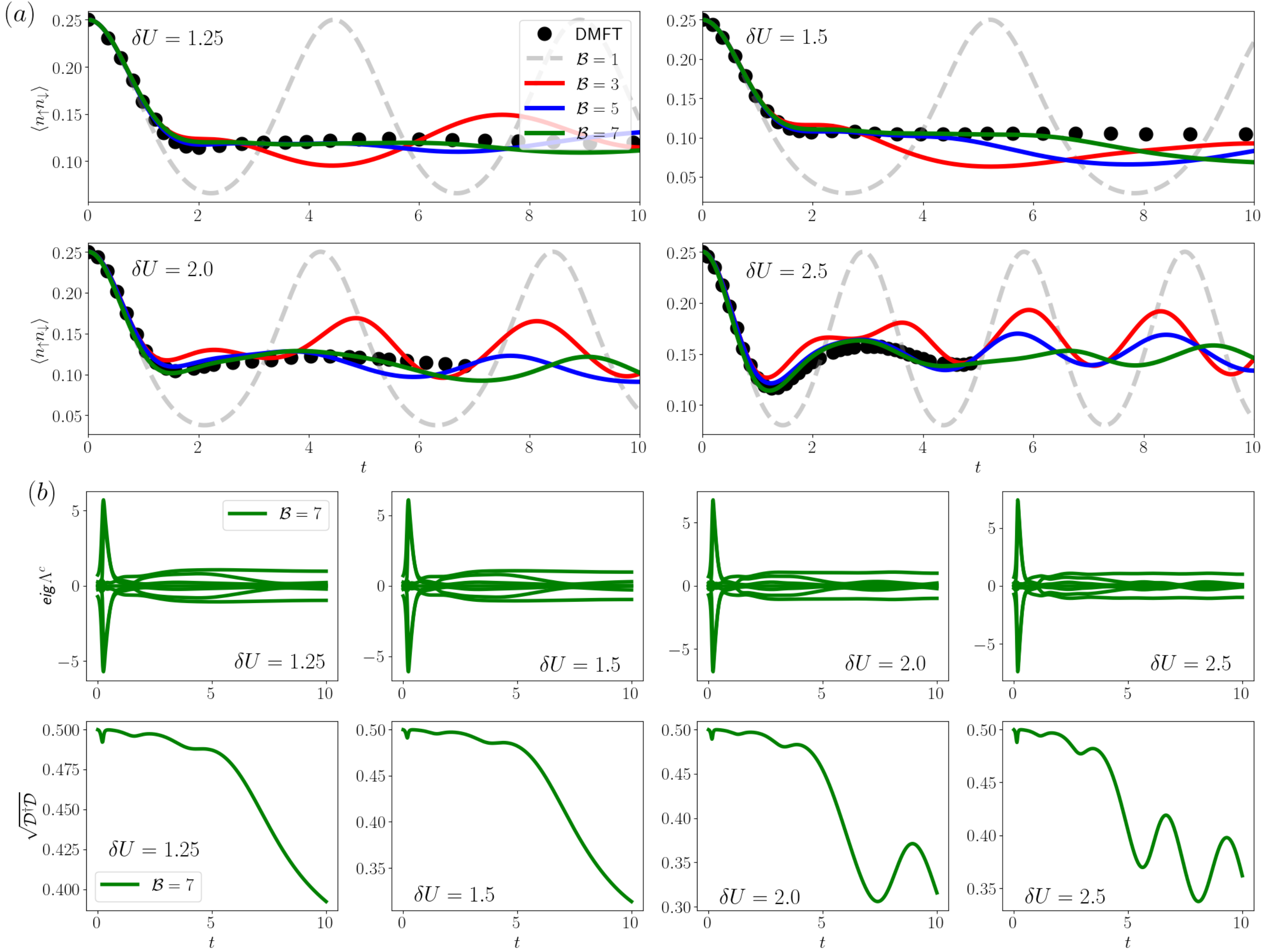}
    \caption{Panel (a): evolution of the double occupancies as a function of time for different values of $\delta U$, in sequence $\delta U=1.25,1.5,2.0,2.5$, respectively. 
     Dotted data shows the DMFT result taken from Ref.~\cite{Eckstein_2009}. Different colors correspond to different values of $\mathcal{B}$ (number of bath sites). 
    Panel (b): top row shows the evolution of the eigenvalues of the $\Lambda^c$ for $\mathcal{B}=7$. Bottom row displays the evolution of $\sqrt{\mathcal{D}^\dagger \mathcal{D}}$ for $\mathcal{B}=7$.}
    \label{fig:evolution_multiplot}
\end{figure}

\end{widetext}

We focus our analysis by considering as initial condition a weakly correlated metal $U_i\simeq 0$
and the final interaction strength is larger than the initial value $U_f>U_i$~\footnote{Our calculations have been performed setting the initial state a very small interaction strength $U_{i}=0.05$. This is because the g-GA variational landscape is degenerate in the limiting case $U=0$ $\forall\, \B\geq 3$ and, therefore, the self-consistency procedure becomes numerically unstable.}. 
Under these circumstances, the standard td-GA dynamics (corresponding to $\B=1$ in our formalism) is characterized by the presence of a dynamical quantum critical point that identifies three different dynamical regimes of weak $U_f<U^{\rm dyn}_c$, intermediate $U_f\sim U^{\rm dyn}_{c}$ and strong $U_f>U^{\rm dyn}_c$ quenches~\cite{SchiroFabrizioPRL}.
Within this framework ($\B=1$) different regimes were identified by computing the period of oscillation of the time-dependent double occupancy $d(t)=\mel{\Phi(t)}{n_\uparrow n_\downarrow}{\Phi(t)}$ following the quench, which is purely monocromatic, see Fig.~\ref{fig:periodo_standard}.

Below we show the td-gGA time evolution of the double occupancy $d(t)$ as a function of $\B$, in comparison with the numerically-exact td-DMFT results of Ref.~\cite{Eckstein_2009}, for different values of $U_f$, spanning all the different dynamical regimes, see Fig.~\ref{fig:evolution_multiplot}a.
In order to better interpret the results we also show the time evolution of the eigenvalues of $\Lambda^c$ and $\D^\dagger\D$, which are both gauge-invariant quantities associated with the dynamics of the EH, see Fig.~\ref{fig:evolution_multiplot}b. This allows us to detect the physical dynamics of the embedding parameters, decoupling it from irrelevant time-dependent gauge transformations.

Our results show that, while within the standard td-GA ($\B =1$) the evolution of $d(t)$ is accurate only at very short times and does not capture the relaxation of the double occupancy at long times which is observed in td-DMFT, for $\B\geq 3$ we develop a clear trend towards a relaxation of the local observables which replaces the oscillations obtained for $\B =1$. The improvement introduced by introducing the auxiliary degrees of freedom approches the td-DMFT dynamics with increasingly high accuracy.
Indeed the td-gGA dynamics follows the DMFT reference on a timescale that increases as we increase $\B$. For instance, we note that for values as small as $\B=7$ the method achieves nearly perfect agreement with td-DMFT for $t \lesssim 6$, for all quenches considered. 
It is interesting to note that the td-gGA dynamics of the double occupancy arises from the time dependent Schr\"odinger equation [Eq.~\eqref{dyn-Phi}], and the parameters of the corresponding EH shown in the middle and right columns of the figure evolve in time even when $d(t)$ appears to be essentially stationary. This is consistent with the general fact that the equilibration arises for local quantities, such as the double occupancy, even though  the quantum dynamics of the many-body electronic function, here encoded in the time evolution of the td-gGA variational parameters, is unitary.

\section{Conclusions}
\label{sec:conclusions}

In this work, we introduced a time-dependent extension of the ghost Gutzwiller approximation (gGA) for the study of correlated electron systems in nonequilibrium situations. We have benchmarked the method for an interaction quench of the half-filled Hubbard model, comparing explicitly with one of the state-of-the-art approaches, td-DMFT.

Our results clearly show that this approach, already for a small number $\B$ of auxiliary parameters,  improves qualitatively on the standard Gutzwiller approximation, since it can describe the relaxation of local observables, and it achieves a remarkable quantitative agreement with td-DMFT for a wide range of model parameters and timescales already for small values of $\B$.

A crucial point is that an accuracy comparable with td-DMFT is thus obtained at a hugely smaller computational cost since the td-gGA requires to solve a non-linear ordinary differential equation as opposed to the integral-differential equation required by td-DMFT. 

%
In particular, all calculations performed in this work have been performed serially on a single CPU, highlighting the method's computational efficiency. 
Utilizing methods such as time-dependent density matrix renormalization group (DMRG)~\cite{Wolf_2014,Bauernfeind_2022} or matrix products states (MPS)~\cite{PhysRevLett.105.050404,Kohn_2021,Thoenniss_2022,Ng_2022} to solve the time-dependent Schr\"odinger equation of the EH would allow us to reduce the computational complexity even further, allowing us to perform calculations with more bath sites, even for multi-orbital strongly correlated systems.
Furthermore, the very fact that td-gGA ultimately reduces to a finite-dimensional first-order non-linear differential equation, allows one to employ, e.g.,  optimal control methods to steer a given dynamical system to desired outcomes.
These observations suggests that the td-gGA has the potential to advance our ability to study the non-equilibrium properties of a variety of systems of great interest, ranging from the general study of  multi-orbital correlated electron materials to quantum devices including  Mott solar cells~\cite{Manousakis_2010,PhysRevB.90.235102,PhysRevApplied.3.064015,Petocchi_2019,Petocchi_2022}, or more in general to any problem which requires a proper treatment of electronic correlations while accessing non-equilibrium properties.

\section{Acknowledgements}

We are in debt with Michele Fabrizio for his insightful contributions at the early stage of the project. 
We also acknowledge discussions with Marco Schir\'o, Carlos Mejuto Zaera and Jan Skolimowski. 
Flatiron Institute is a division of the Simons Foundation. 
This work was supported by a grant from the Simons Foundation (1030691, NL). 
NL gratefully acknowledges funding from the Novo Nordisk Foundation
through the Exploratory Interdisciplinary Synergy Programme project
NNF19OC0057790. M.C. acknowledges financial support from MUR via PNRR MUR project PE0000023-NQSTI, PNNR National Center for HPC, Big Data, and Quantum Computing (grant No. CN00000013), PRIN 2020 q-LIMA (Protocol  Number 2020JLZ52N),and PRIN 2017 CEnTral (Protocol Number 20172H2SC4). DG  acknowledges the hospitality of the Aspen Center for Physics where this work was finalized.



%

\end{document}